\documentclass[doublecol]{epl2} 

\usepackage[english]{babel}
\usepackage{amsmath,amssymb,amsfonts}
\usepackage{braket}

\renewcommand{\Re}{\mathop{\mathrm{Re}}}

\usepackage{silence}
\usepackage[colorlinks=True,linkcolor=blue,citecolor=blue,urlcolor=blue]{hyperref}

\title{The non-Hermitian skin effect: A perspective}

\author{Julius T. Gohsrich\inst{1,2,a} \and Ayan Banerjee\inst{1,b} \and Flore K. Kunst\inst{1,2,c}}
\shortauthor{J.T. Gohsrich, A. Banerjee, and F.K. Kunst}

\institute{                    
  \inst{1} Max Planck Institute for the Science of Light, Staudtstr. 2, 91058 Erlangen, Germany\\
  \inst{2} Department of Physics, Friedrich-Alexander Universität Erlangen-Nürnberg, Staudtstr. 7, 91058 Erlangen, Germany\\
  \inst{a} \href{mailto:julius.gohsrich@mpl.mpg.de}{\emph{julius.gohsrich@mpl.mpg.de}} \qquad \qquad \inst{b} \href{mailto:ayan.banerjee@mpl.mpg.de}{\emph{ayan.banerjee@mpl.mpg.de}} \qquad \qquad \inst{c} \href{mailto:flore.kunst@mpl.mpg.de}{\emph{flore.kunst@mpl.mpg.de}}
}

\abstract{ The non-Hermitian (NH) skin effect is a truly NH feature, which manifests itself as an accumulation of states, known as skin states, on the boundaries of a system. In this perspective, we discuss several aspects of the NH skin effect focusing on the most interesting facets of this phenomenon. Beyond reviewing necessary requirements to see the NH skin effect, we discuss the NH skin effect as a topological effect that can be seen as a manifestation of a truly NH bulk-boundary correspondence, stemming from the spectral topology, and show how skin states can be distinguished from topological boundary states. As most theoretical work has focused on studying the NH skin effect in one-dimensional non-interacting systems, recent developments of studying this effect in higher dimensions as well as in many-body systems are highlighted. Lastly, experimental signatures and applications are discussed, and an outlook is provided.}

\begin{document}

\maketitle

\section{Introduction}

Non-Hermitian (NH) topology has established itself as a new exciting research domain over the last decade~\cite{bergholtz_exceptional_2021,ashida_non-hermitian_2020,okuma_non-hermitian_2023}. Originally motivated from the fact that parity-time symmetric systems may have a completely real eigenspectrum~\cite{bender_real_1998,bender_making_2007}, and the subsequent surge in optical experiments~\cite{el-ganainy_non-hermitian_2018,miri_exceptional_2019,ozdemir_paritytime_2019, weidemann_topological_2020,xiao_non-hermitian_2020}, non-Hermiticity is increasingly studied in a wide variety of fields in physics ranging from mechanical~\cite{ghatak_observation_2020}, electrical~\cite{helbig_generalized_2020}, acoustic~\cite{zhang_acoustic_2021} and open quantum systems~\cite{song_non-hermitian_2019-1,yang_liouvillian_2022} to quantum scattering~\cite{franca_non-hermitian_2022}, strongly correlated phases of matter~\cite{nakamura_non-hermitian_2006,nakagawa_dynamical_2020, lee_many-body_2020,liu_non-hermitian_2020,alsallom_fate_2022,yoshida_non-hermitian_2024,shimomura_general_2024,liang_dynamic_2022} and non-conservative biological systems~\cite{nelson_non-hermitian_1998,lubensky_pulling_2000}. In the field of NH topology, non-Hermiticity is studied through the lens of condensed matter physics uncovering a remarkable enrichment of topological phenomena~\cite{bergholtz_exceptional_2021}.

\begin{figure*}[ht]	
\includegraphics[width=\linewidth]{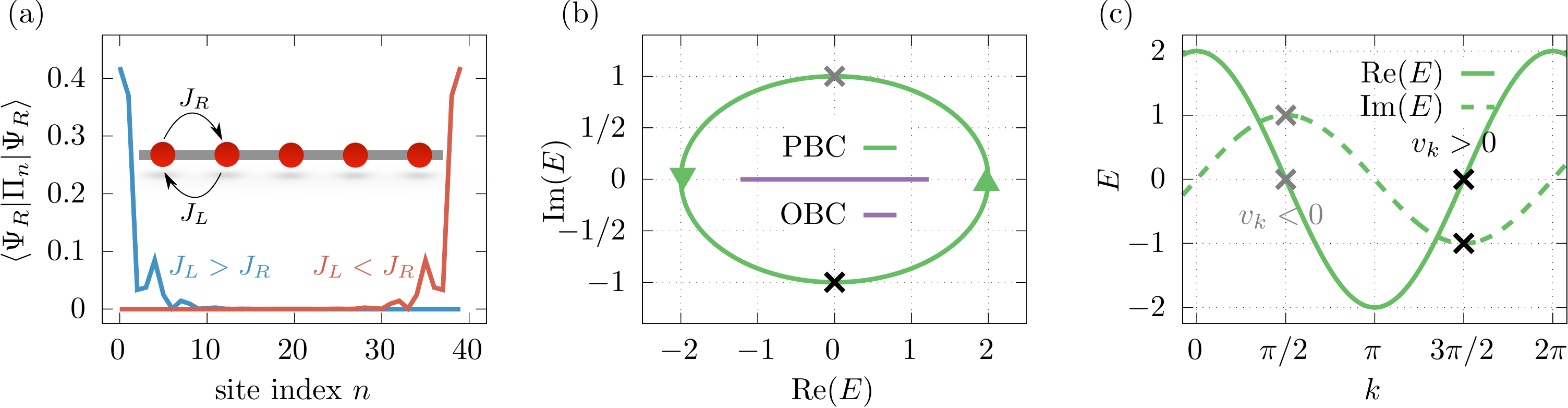}
\caption{Hatano-Nelson model and its features. (a) Schematic illustration of the model showing the localization of bulk modes under OBCs on the right (left) when $|J_L|<|J_R|$ ($>|J_R|$). (b) Under PBCs, the eigenvalues trace an ellipse in the complex plane (green), and the eigenstates are extended. In contrast, the OBC spectrum (purple) drastically deviates from the PBC spectrum exhibiting purely real eigenvalues and localized eigenmodes manifesting the NH skin effect. (c) The real (solid) and imaginary (dashed) parts of the PBC energies as a function of $k$. The gray and black crosses in (b) and (c) indicate the left and right movers in the dispersion, respectively. $J_L = 3/2$ and $J_R = 1/2$ in (b) and (c).}
\label{HN}	
\end{figure*}

At the core of this enrichment is the fact that not only the eigenvectors, but also the eigenvalues may have non-trivial topological features. Indeed, the spectrum now lives on the complex plane, such that eigenvalues may wind in a non-trivial manner captured by a spectral invariant~\cite{leykam_edge_2017,shen_topological_2018,gong_topological_2018, okuma_topological_2020,zhang_universal_2022} as illustrated in Fig.~\ref{HN}. It has been shown that topologically non-trivial spectral features under periodic boundary conditions (PBCs) result in a macroscopic accumulation of states on the boundary of a lattice model under open boundary conditions (OBCs)~\cite{yao_edge_2018,borgnia_non-hermitian_2020,okuma_topological_2020, zhang_correspondence_2020,zhang_universal_2022, koekenbier_transfer_2024-1} establishing a truly \emph{non-Hermitian bulk-boundary correspondence}. These exponentially localized states are the so-called NH skin states, and the phenomenon overall is known as the \emph{non-Hermitian skin effect}, a term coined in Ref.~\citenum{yao_edge_2018}, which is the central subject of this perspective.

\section{The NH skin effect: The basics}
We start this perspective with a paradigmatic one-dimensional (1D) example, to discuss several general properties we believe are well-known, yet important to understand the NH skin effect in 1D systems.

\subsection{The Hatano-Nelson model} The Hatano-Nelson model was first proposed by Hatano and Nelson to study localization transitions in the context of superconductivity~\cite{hatano_localization_1996}, and its Hamiltonian, shown in the inset of Fig.~\ref{HN}(a), reads
\begin{equation}
H = \sum_n (J_L c^\dagger_{n} c_{n+1} + J_R c^\dagger_{n+1} c_{n}), \label{eq:HN_Ham}
\end{equation}
with $J_L \, (J_R) \in \mathbb{R}$ the nearest-neighbor (NN) hopping parameter to the left (right), and $c^\dagger_{n}$ ($c_{n}$) creating (annihilating) an excitation on site $n$. This model is NH when $J_L \neq J_R$.
As a 1D single-band model, the energy under PBCs $E(k)$ corresponds to the Bloch Hamiltonian $H(k)$, and reads $E(k) = (J_L + J_R) \cos k + i (J_L - J_R) \sin k$. As seen in Fig.~\ref{HN}(b), $E(k)$ forms an ellipse and winds around the origin in the complex energy plane as a function of $k$ in the (counter)clockwise direction, when $|J_L|<|J_R|$ ($>|J_R|$). As such, it is possible to define a \emph{spectral} winding number $w$ as~\cite{leykam_edge_2017,gong_topological_2018,shen_topological_2018}
\begin{equation}
w= \frac{1}{2 \pi i} \int_{-\pi}^\pi \mathrm{d} k \, \partial_k \ln E(k) = \begin{cases}
    +1, & |J_L|>|J_R|, \\
    -1, & |J_L|<|J_R|.
\end{cases} \label{eq:winding_number_HN}
\end{equation}
The winding number can only change when $E(k) = 0$ for some $k$, i.e., when $|J_L| = |J_R|$. Turning to OBCs, it is straightforward to see from the Hamiltonian in Eq.~\eqref{eq:HN_Ham}, that all eigenstates satisfying $H \ket{\Psi_R} = E \ket{\Psi_R}$ are exponentially localized on the boundary on the right (left) when $|J_L|<|J_R|$ ($>|J_R|$) as shown in Fig.~\ref{HN}(a). This phenomenon of a piling up of a macroscopic number of states on the boundary is the NH skin effect.

We find that there is an intuitive and enlightening way to understand the connection between the sign of the spectral winding number and the boundary to which the skin states localize. When the winding number is negative (positive), the eigenvalues wind in the (counter)clockwise direction. As the PBC spectrum of the Hatano-Nelson model is centered around zero energy, half of the spectrum is in the positive imaginary plane. That means that for a positive winding number, the right (left) movers with group velocity $v_k = \Re [\partial_k E(k)] >0 $ ($<0$) have negative (positive) imaginary energy, as illustrated in Figs.~\ref{HN}(b,c). In the long-time limit, the left movers thus dominate the behavior of the system, and one sees that all NH skin states pile up on the left boundary of the system. The reverse holds for negative winding number. A similar argument is presented in Ref.~\citenum{lee_topological_2019}.

\subsection{Localization of skin states} When considering eigenstates of a NH system, one should take care of the appropriate biorthogonalization~\cite{brody_biorthogonal_2013}: A NH system does not only have right eigenstates $\ket{\Psi_R}$, but also distinct left eigenvectors satisfying $\bra{\Psi_L} H = E \bra{\Psi_L}$, equivalent to $H^\dagger \ket{\Psi_L} = E^* \ket{\Psi_L}$. By studying the left and right eigenstates separately one can observe the skin effect. Indeed, if $\ket{\Psi_R}$ is localized on one boundary, the corresponding $\bra{\Psi_L}$ is localized to the opposite boundary~\cite{bottcher_spectral_2005,yokomizo_non-bloch_2019,koekenbier_transfer_2024-1}. As a consequence, the \emph{biorthogonal localization} of the right and left eigenstate quantified by $\braket{\Psi_L|\Pi_n|\Psi_R}$, where $\Pi_n = c_n^\dagger \ket{0} \bra{0} c_n$ is the projector onto each site $n$, has weight throughout the system. Thus, skin states have a bulk-state-like fingerprint in the biorthogonal picture, albeit with potentially negative or complex weights.

\subsection{Generalized Brillouin zone}

Next, we point out the intimate link between the NH skin effect and the so-called generalized Brillouin zone (GBZ). The theory of the GBZ was first developed by Yao and Wang in Ref.~\citenum{yao_edge_2018}, and later expanded in Refs.~\citenum{song_non-hermitian_2019} and \citenum{yokomizo_non-bloch_2019}. The GBZ is a closed curve $\mathcal{C}$ solving the characteristic equation of the non-Bloch Hamiltonian $H(\beta)$, which is the analytic continuation of the Bloch Hamiltonian. The GBZ contains crucial information about the system under OBCs: The OBC spectrum is given by $H(\beta)$ with $\beta \in \mathcal{C}$, and one can determine spatial decay rates for the NH skin modes, which are related to $\beta$, in the continuum limit. Indeed, for any skin mode, one finds its associated $\beta$, and if $|\beta|<1$ $(>1)$, or in other words if $\mathcal{C}$ is inside (outside) of the unit circle~\cite{song_non-hermitian_2019}, the skin state localize to the left (right) boundary. For $|\beta|=1$, the state is delocalized as the Bloch states, and the corresponding point is called a Bloch point~\cite{song_non-hermitian_2019}.

\subsection{Exceptional points} 

Interestingly, the NH skin effect is accompanied by the appearance of exceptional points (EPs) with an order scaling with system size under OBCs~\cite{martinez_alvarez_non-hermitian_2018,kunst_non-hermitian_2019}. EPs are degeneracies, which abundantly appear in NH matrices~\cite{berry_physics_2004,budich_symmetry-protected_2019}, at which not only the eigenvalues but also the eigenvectors coalesce, i.e., the geometric multiplicity is smaller than the algebraic multiplicity~\cite{kato_perturbation_1995, heiss_physics_2012}. At EPs, the NH matrix features Jordan blocks, whose dimension corresponds to the order of the EP. Close to an EP, eigenvectors start to overlap until they finally coalesce when reaching the EP. We indeed see such a high-order EP for the Hatano-Nelson model, where in the extreme limit $J_R \, (J_L) = 0$, all states coalesce onto a single one with weight only on the leftmost (rightmost) site. 

\subsection{Sensitivity to boundary conditions}

Many questions often arise in relation to the thermodynamic limit in NH systems. Here we highlight how subtle this matter is by studying the sensitivity to boundary conditions in the Hatano-Nelson model.
As is clearly visible in the spectrum in Fig.~\ref{HN}(b), it strongly depends on the boundary conditions. In Ref.~\citenum{xiong_why_2018}, it is shown that by introducing coupling terms between the ends of 1D chains, one finds that the system crosses several EPs upon tuning tune between OBCs and PBCs~\cite{kunst_biorthogonal_2018,koch_bulk-boundary_2020,roccati_non-hermitian_2021}, emphasizing a topological distinction between the two cases. Refs.~\citenum{kunst_biorthogonal_2018} and \citenum{koch_bulk-boundary_2020} show that the cross-over from OBCs to PBCs, i.e., from having skin states to Bloch states, happens for exponentially small couplings. While it thus may seem that the NH skin effect is not a robust feature, it is shown in Ref.~\citenum{koch_bulk-boundary_2020} that physically relevant locality constraints exponentially suppress this kind of coupling. Also, further modifications to the boundary conditions result in new localization phenomena and transitions between different behaviors~\cite{li_impurity_2021}.

\subsection{Beyond Hamiltonians}

While usually phrased in terms of Hamiltonians, we note that the skin effect is a property of NH matrices or operators. Thus, any physical system represented by such may exhibit the skin effect. Examples include the conductance matrix, which is the appropriate object to consider when handling topoelectric circuits \cite{helbig_generalized_2020}, the damping matrix in open quantum systems~\cite{song_non-hermitian_2019-1,yang_liouvillian_2022}, and the scattering matrix~\cite{franca_non-hermitian_2022,brunelli_restoration_2023}.

\section{Non-normality and non-reciprocity as requirement of the NH skin effect}
We want to point out that not all NH matrices feature the NH skin effect, and a necessary criterion needs to be satisfied, namely, the NH matrix needs to be \emph{non-normal}~\cite{bergholtz_exceptional_2021}, i.e., $[H,H^\dagger] \neq 0$. By writing $H = H_\text{H} + i H_\text{A}$, with $H_\text{H} = H_\text{H}^\dagger$ the Hermitian part and $i H_\text{A} = -(iH_\text{A})^\dagger$ the anti-Hermitian part of $H$, non-normality is equivalent to $[H_\text{H}, H_\text{A}] \neq 0$. If $H_\text{H}$ and $H_\text{A}$ would commute, $H$ would share an orthogonal eigenbasis with $H_\text{H}$ and $H_\text{A}$, which prohibits the existence of skin states.

Additionally, a system needs to be \emph{non-reciprocal} to show the NH skin effect~\cite{brunelli_restoration_2023}. This can either be checked by analyzing if the systems Green's function, $\chi (E) = -i (E \,\mathbb{I} - H)^{-1}$, satisfies $|\chi (E)| \neq |\chi (E)|^T$, where the modulus of a matrix means the absolute value of each matrix element, or by analyzing the systems' singular value decomposition (SVD)~\cite{brunelli_restoration_2023}. In the scattering context discussed in Ref.~\citenum{brunelli_restoration_2023}, the requirement of non-reciprocity is equivalent to the requirement that the systems scattering matrix $S$ satisfies $|S| \neq |S|^T$. Physically, this means that the response of the system is not invariant under exchanging input and output~\cite{jalas_what_2013,caloz_electromagnetic_2018-1}. While non-normality and non-reciprocity are equivalent for single-band NN models, e.g., the aforementioned Hatano-Nelson model, this is in general not true for single-band systems including longer-range hoppings and multi-band systems~\cite{brunelli_restoration_2023}.

We want to emphasize that the presented definition of non-reciprocity is different from the notion of non-reciprocal hoppings, i.e, hoppings whose magnitude depends on the hopping direction, which we refer to as \emph{asymmetric} hoppings. It is namely possible to observe the skin effect \emph{without} asymmetric hoppings: While the frequently studied anisotropic Su-Schrieffer-Heeger (SSH) model in Eq.~\eqref{eq:nh_SSH} below features asymmetric hoppings, it is unitarily equivalent to the NH Creutz ladder~\cite{bergholtz_exceptional_2021} introduced by Lee as a first example for the breakdown of the conventional bulk-boundary correspondence (cBBC)~\cite{lee_anomalous_2016,bergholtz_exceptional_2021}. The NH Creutz ladder does not exhibit asymmetric hoppings, but instead features on-site gain and loss as well as phases on the hoppings. Nevertheless, this model is non-reciprocal in the aforementioned sense and shows the NH skin effect.

\begin{figure*}[t]	
\includegraphics[width=\linewidth]{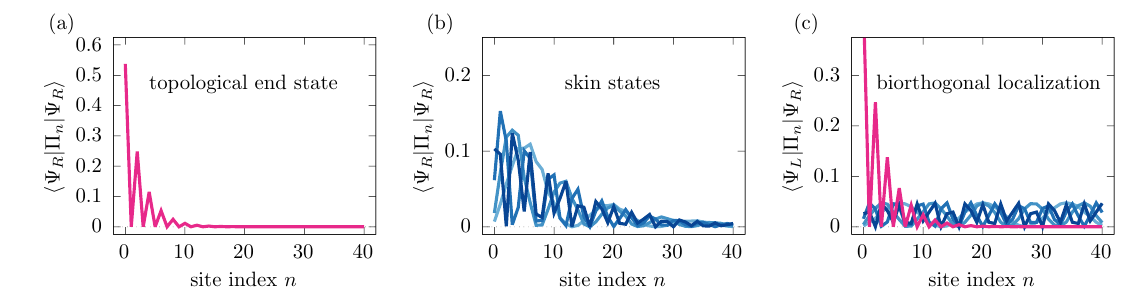}
\caption{Topological end state and skin state localization. Normalized right eigenstate of (a) topological end state in pink and (b) skin states in different shades of blue for the model in Eq.~\eqref{eq:nh_SSH}. The end state is localized on a single sublattice whereas the skin modes are distributed across both sublattices. (c) The biorthogonal product of the right and left states shows that the end state remains localized while the skin states are delocalized. We fix $t_1=3/2, t_2=2,$ $\gamma=0.14$, and $N=41$.}
\label{distict-edge-skin}	
\end{figure*}

\section{Spectral features, topological protection and the NH bulk-boundary correspondence}

If a NH matrix is both non-normal and non-reciprocal, its PBC spectrum displays a so-called point gap~\cite{brunelli_restoration_2023}. A point gap is defined as there being a base point $E_B$, which is not crossed by the complex-energy bands, and crossing this point defines a gap closing transition~\cite{gong_topological_2018}. Indeed, we see in the example of the Hatano-Nelson model that the PBC spectrum has a point gap, cf. Fig.~\ref{HN}(b), and the gap closes as discussed at $|J_L|=|J_R|$. It has been shown that for the NH skin effect to occur in 1D systems, the PBC spectrum must feature such point gaps~\cite{okuma_topological_2020, zhang_correspondence_2020,nakamura_bulk-boundary_2024,koekenbier_transfer_2024-1}.
By extension, this means that it is possible to find a non-zero spectral topological invariant for these systems. For example, the spectral winding number in Eq.~\eqref{eq:winding_number_HN} for arbitrary periodic 1D systems described by the Bloch Hamiltonian $H(k)$, can be generalized to the following form~\cite{gong_topological_2018}:
\begin{equation}
    w(E_B) = \frac{1}{2 \pi i} \int_{-\pi}^\pi \mathrm{d} k \, \partial_k \ln\{\det[H(k)] - E_B\}.
\end{equation}
We note that if one is interested in the state localization of a specific skin state, one has to select its energy as reference energy $E_B$. While it is quite intuitive that all eigenstates are skewed towards the right (left) when $|J_L|<|J_R|$ ($>|J_R|$) in the Hatano-Nelson model, the spectral winding number is essential to understand the competition between different hopping strengths and ranges in more general models.

The skin effect as a topological effect is robust against weak perturbations of the system. For example, the Hatano-Nelson model, cf. Eq.~\eqref{eq:HN_Ham}, is robust against weak onsite disorder~\cite{hatano_localization_1996,gong_topological_2018} and additional perturbations to the hoppings~\cite{claes_skin_2021,wanjura_correspondence_2021}, revealing a stability of the skin states to these disturbances.

These insights regarding the link between a spectral invariant defined under PBCs, and the NH skin effect then lead to the establishment of a \emph{new, truly non-Hermitian bulk-boundary correspondence}: Having point gaps in the PBC spectrum of 1D models quantified by a non-trivial spectral invariant implies the piling up of states on the boundary of the system under OBCs~\cite{okuma_topological_2020, zhang_correspondence_2020, borgnia_non-hermitian_2020, nakamura_bulk-boundary_2024 ,koekenbier_transfer_2024-1}. 

\subsection{Beyond spectral winding numbers: Symmetries}
There are models, which feature a NH skin effect, while having a zero spectral winding number. In these cases, it is possible to find alternative spectral invariants. For example, the presence of the so-called time-reversal dagger symmetry (TRS$^\dagger$)~\cite{kawabata_symmetry_2019} with the symmetry operator squaring to minus one results in doubly degenerate loops in the PBC spectrum with opposite winding numbers. As a result, half of the skin states accumulate on one boundary and the other half on the other. The spectral invariant is a $\mathbb{Z}_2$ invariant, and this type of skin effect is referred to as the $\mathbb{Z}_2$-skin effect~\cite{okuma_topological_2020}. In general, symmetries play an important role in the context of the NH skin effect. Indeed, the presence of certain symmetries prohibits the appearance of the NH skin effect. These symmetries include parity-time symmetry~\cite{kunst_non-hermitian_2019} and pseudo-Hermitian symmetry~\cite{kawabata_symmetry_2019} in the unbroken phase for any-dimensional systems as well as parity symmetry and TRS$^\dagger$ with the symmetry operator squaring to one for 1D systems~\cite{kawabata_symmetry_2019}.

\section{Topological boundary states and the NH skin states}
Next we turn to the coexistence of skin states and topological boundary states, and discuss how to distinguish them, which we believe is important to understand. Previously, we saw that the NH skin effect is protected by spectral topology, specifically the point gap topology unique to NH systems. In contrast, topological boundary states arise from wavefunction topology. If a model features both NH skin states and topological boundary states, it is a priori not obvious how to distinguish these different types of states.

To illustrate this distinction, we focus on the anisotropic SSH model described by the Hamiltonian~\cite{kunst_biorthogonal_2018, yao_edge_2018}
\begin{multline}
    H = \sum_{n} \big[ (t_1 + \gamma) c_{A,n}^\dagger c_{B,n} + (t_1 - \gamma) c_{B,n}^\dagger c_{A,n} \\[-0.5em]
    + t_2 (c_{A,n+1}^\dagger c_{B,n} + c_{B,n+1}^\dagger c_{A,n}) \big], \label{eq:nh_SSH}
\end{multline}
where $c_{\alpha,n}^\dagger$ $(c_{\alpha,n})$ creates (annihilates) an excitation on sublattice $\alpha =A, B$ in unit cell $n$, $t_1$ ($t_2$) is the NN hopping parameters inside (between) unit cells, and $\gamma$ makes the system NH by changing the magnitude of the hopping to the right with respect to the hopping to the left, i.e., it makes the hoppings asymmetric. This model exhibits sublattice symmetry, which allows for the explicit construction of the topological end states via a destructive interference argument~\cite{kunst_biorthogonal_2018}. Such end states are also exponentially localized, however, they only have weight on one of the sublattices, as shown in Fig.~\ref{distict-edge-skin}(a) for a right eigenstate localized to the left boundary. Due to the anisotropy factor $\gamma$, all the other states pile up on one of the boundaries, cf. Fig.~\ref{distict-edge-skin}(b), thus displaying the NH skin effect. Let us now answer the question as to how to distinguish the topological boundary states from the skin states.

The answer to this question lies in the \emph{biorthogonal properties} of these states. While the skin states behave as bulk states when studying their biorthogonal localization, as we already discussed before, topological boundary states will remain localized to the boundary, as shown in Fig.~\ref{distict-edge-skin}(c). We note that situations may also arise in which the right eigenstate of a boundary-state solution is localized to the opposite boundary as compared to the left eigenstate, as also mentioned in Ref.~\citenum{budich_non-hermitian_2020} in the context of topological sensors discussed below. In this case, the biorthogonal product of the right and left topological boundary state will still show a stronger signature to one of the boundaries, while preserving its general profile.

We remark that this model breaks the cBBC~\cite{kunst_biorthogonal_2018, yao_edge_2018}, where a topological invariant defined from the Bloch Hamiltonian predicts the existence of modes on the boundaries. The previously mentioned GBZ approach provides a route towards restoring the cBBC~\cite{yao_edge_2018, song_non-hermitian_2019, yokomizo_non-bloch_2019}, whereas an alternative approach based on the biorthogonal properties of the boundary modes was proposed in Ref.~\citenum{kunst_biorthogonal_2018}. This breakdown of the cBBC, first noticed in Ref.~\citenum{lee_anomalous_2016}, is a common feature of topological models featuring the NH skin effect~\cite{kunst_non-hermitian_2019}, and can be intuitively understood from the fact that a discrepancy between the PBC and OBC spectra means gap closings indicating topological phase transitions generally occur for different parameter values.

\section{Skin effect in higher dimensions}

So far we have only considered the appearance of skin states in 1D systems. In higher dimension, there is not yet an all-encompassing theory, which we believe to be an important future direction of study. Efforts in this area have been made very recently with the development of the amoeba formalism~\cite{wang_amoeba_2024}, the dual Newton polygon formalism~\cite{brieskorn_plane_2012}, and the tropical geometric framework~\cite{maclagan_introduction_2015} to characterize the NH skin effect~\cite{jaiswal_characterizing_2023,banerjee_tropical_2023, wang_amoeba_2024, wang_constraints_2024,hu_topological_2025}.

\begin{figure}[t]	
\includegraphics[width=\linewidth]{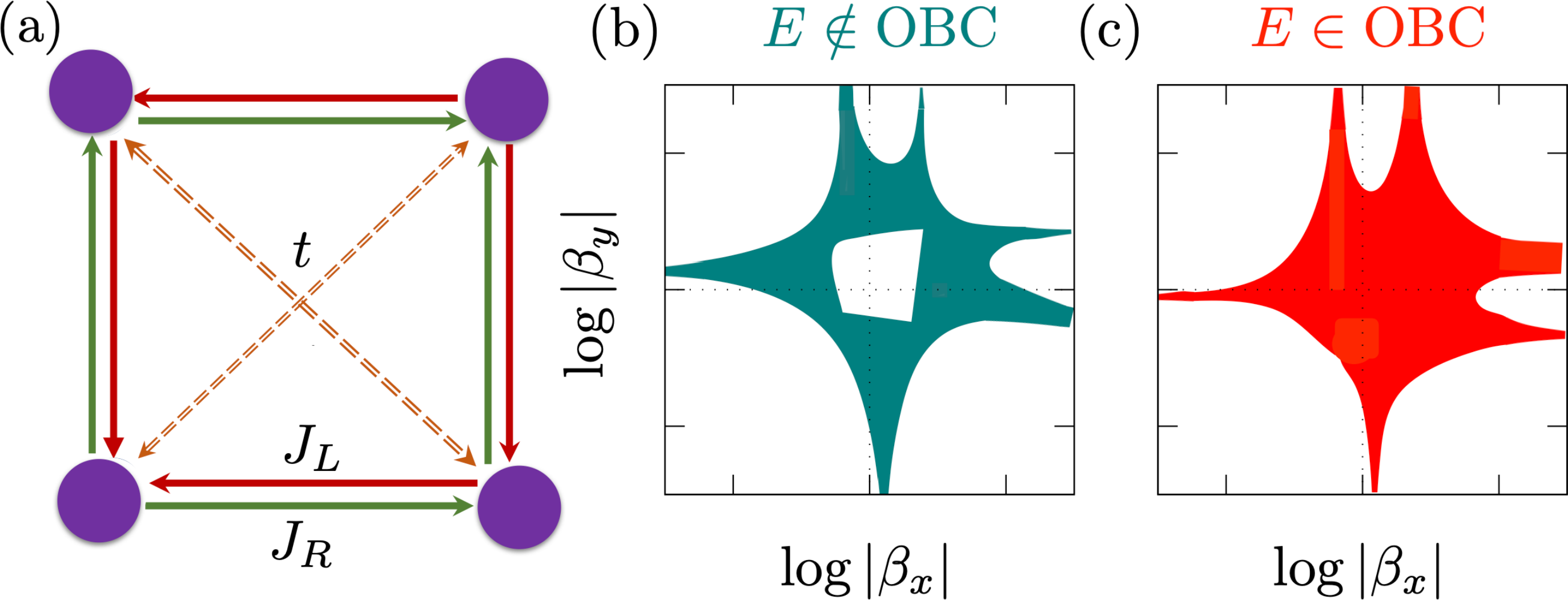}
\caption{Amoeba formulation in 2D. (a) Illustration of the 2D single-band model with asymmetric hoppings, Eq.~\eqref{eq:2D_example}. (b,c) The plots depict 2D amoeba, where the points $(\log|\beta_x|, \log|\beta_y|)$ satisfy the characteristic equation Eq.~\eqref{eq:det}. A hole appears in the amoeba for energies outside the OBC spectrum in (b), while in (c), no hole is present for energies within the OBC spectrum. We set $J_R=4/5, J_L=6/5$, and $t=1/2$.}
\label{amoeba}
\end{figure}

To highlight these new developments, let us discuss the amoeba formalism with a simple example of a single-band model in two dimensions (2D)~\cite{wang_amoeba_2024}, shown in Fig.~\ref{amoeba}(a). Its Hamiltonian reads
\begin{align}
    \label{eq:2D_example}
    H &=\sum_{n,m}\left[J_L \left(c^{\dagger}_{n,m}c_{n+1,m}+
    c^{\dagger}_{n,m+1}c_{n,m}\right) \right. \notag \\[-0.5em] 
    &\quad \ +J_R\left(c^{\dagger}_{n+1,m}c_{n,m} +c^{\dagger}_{n,m}c_{n,m+1}\right)  \notag\\
    &\quad \ + t \left. \left(c^{\dagger}_{n,m}c_{n+1,m+1}+ c^{\dagger}_{n,m+1}c_{n+1,m} +\text{H.c.}\right)\right],
\end{align}
where $c^{\dagger}_{n,m}$ ($c_{n,m}$) creates (annihilates) an excitation on site $(n,m)$ in the 2D lattice, and $J_L$ and $J_R$ ($t$) are the (next-)nearest neighbor hoppings. To determine the Bloch Hamiltonian, one Fourier transforms to introduce the two Bloch momenta $k_x$ and $k_y$. As in the GBZ theory, one analytically continues the Bloch momenta, i.e., one replaces the plane waves $(e^{ik_x},e^{ik_y})$ by the complex functions $(\beta_x,\beta_y)$ to find the non-Bloch Hamiltonian $H(\beta_x,\beta_y)$. Then, the formalism relies on the principle that all algebraic-geometric information has to be encoded in its characteristic equation
\begin{equation}
\label{eq:det}
    \det[E-H(\beta_x,\beta_y)]=0.
\end{equation}
An amoeba is a region in $\mathbb{R}^d$ solving this equation ($d=2$ in our example), resembling its biological prototype~\cite{gelfand_discriminants_1994}, having holes and narrowing tentacles extending to infinity.

In the example, we can construct the amoebas without solving this polynomial fully for $\beta_x$ and $\beta_y$. Expressing $\beta_y(E)$ as a function of $\beta_x(E)$ as $\beta_{y,A}(\beta_x,E)$, where $A=1,2$ labels the different solutions in $\beta_x$, we can plot $\log|\beta_x(E)|$ against $\log|\beta_{y,A}(E)|$ for any $A$ and $E$ to get amoebas, depicted in Fig.~\ref{amoeba}(b,c). If \emph{any} amoeba for \emph{any} $A$ contains \emph{no} central hole, $E$ belongs to the OBC spectrum, and one can construct the full OBC spectrum in this fashion. The associated Ronkin function then provides insight into the skin state localization lengths~\cite{wang_amoeba_2024}. In 1D, the construction of the amoeba reduces to the construction of the GBZ.

We also remark that the NH skin effect may also appear on higher-order boundaries in higher-dimensional models~\cite{zhang_universal_2022, lee_hybrid_2019,edvardsson_non-hermitian_2019,kawabata_higher-order_2020}, and has also been observed in higher-dimensional non-periodic systems such as quasicrystals and amorphous networks as well as fractals lattices~\cite{manna_inner_2023}.

\section{Skin effect in many-body systems}

Another active topic of current research we find important and want to highlight is the NH skin effect in the context of many-body physics. Here, fermionic and bosonic systems show significant differences. For example, the fermionic case is quite intricate due to the interplay of non-orthogonal eigenstructures and Pauli's exclusion principle~\cite{alsallom_fate_2022}. Also, fermionic repulsion significantly alters the occupied orbitals, ensuring that no more than one fermion, or one hard-core boson, can occupy each physical site. Consequently, the exponential localization of all fermions at a boundary is impossible~\cite{lee_many-body_2020,liu_non-hermitian_2020}. In Ref.~\citenum{alsallom_fate_2022}, it is shown that the many-body skin effect results from an imbalance in the density distribution rather than just a sum of exponential orbitals due to the exclusion principle. The degree of asymmetry in this distribution quantifies the exponential localization of the skin modes as the system size increases. In Ref.~\citenum{shimomura_general_2024}, a general criterion for the appearance of the NH skin effect in many-body systems, known as the Fock-space skin effect, is discussed. Beyond the Pauli principle, interactions enrich the phenomenology of the skin effect. For instance, attractive interactions may cause clustering and thus localization, while repulsive interactions promote delocalization. In contrast to fermionic models, nonreciprocal bosonic models with interacting bosons lead to particle accumulation at the edges, forming a skin superfluid state along with distinct Mott-insulating regimes~\cite{zhang_skin_2020}. Ref.~\citenum{yoshida_non-hermitian_2024} introduces the NH Mott skin effect in a bosonic chain with spin degrees of freedom, emphasizing the interaction between strong correlations and NH point-gap topology.

Turning to the properties of these systems, it has been shown that the skin effect suppresses entanglement propagation hindering thermalization~\cite{abanin_colloquium_2019}. This leads to an area law for the entanglement entropy of the non-equilibrium steady state~\cite{kawabata_entanglement_2023}, in contrast to a volume law in Hermitian systems. More interesting effects are predicted in the many-body context, e.g., skin clustering leading to Hilbert space fragmentation~\cite{shen_non-hermitian_2022}, many-body localization~\cite{wang_non-hermitian_2023}, multifractality~\cite{hamanaka_multifractality_2025}, and multipole skin effect~\cite{gliozzi_many-body_2024}.

\section{Experimental signatures} The NH skin effect has been observed in a plethora of classical experimental platforms, such as in coupled fiber loops~\cite{weidemann_topological_2020}, mechanical metamaterials~\cite{ghatak_observation_2020}, topoelectric circuits~\cite{helbig_generalized_2020}, quantum walks~\cite{xiao_non-hermitian_2020}, acoustic metamaterials \cite{zhang_acoustic_2021}, nano-optomechanical networks~\cite{slim_optomechanical_2024}, and thermal diffusion~\cite{liu_observation_2024}. A particular challenge to observe the NH skin effect is that one has to implement either asymmetric hoppings or complex phases with on-site losses. In this context, we want to highlight the setup in Ref.~\citenum{gao_two-dimensional_2023}, a 2D photonic metamaterial on a chip, which allows to dynamically reconfigure imaginary gauge fields, corresponding to complex phases on the hoppings. Thus, one is able to manipulate the skin effect on the fly in a compact, versatile and scalable platform, promising future applications.

Furthermore, signatures of the skin effect have also been seen in quantum setups such as in cold atoms~\cite{liang_dynamic_2022} and in a multi-terminal quantum Hall device~\cite{ochkan_non-hermitian_2024}. One of such signatures is called self-acceleration, an early-time effect in the dynamics of the system~\cite{longhi_non-hermitian_2022}, was observed in a quantum walk setup~\cite{xue_self_2024}.

While the skin effect can only be directly observed by measuring the right or left eigenstates separately, it is possible to reconstruct them by measuring the system's response. This was done in Ref.~\citenum{zhong_experimentally_2025} in the model discussed in the context of the amoeba formulation, cf. Eq.~\eqref{eq:2D_example}. Measuring the system's response, the Green's function was determined and from it the full complex spectrum as well as the left and right eigenstates. In general, it has been shown that there are observables that crucially depend on the interplay between both flavors of eigenstates, such as the state skewness in the example above, or the system's sensitivity against perturbations as pointed out in this important paper by Schomerus~\cite{schomerus_nonreciprocal_2020}.

\section{Applications}

Beyond observing the existence and effects of skin states, an important development we want to highlight is the harnessing of this effect for real world applications. For example, an experiment using fiber loops proposes a topological funnel for light~\cite{weidemann_topological_2020}. In this setup, two anisotropic 1D chains with opposite winding numbers are coupled to each other so that the skin effect in both chains acts as a funnel resulting in an accumulation of all excitations at the interface. Another proposal to employ the NH skin effect is in the context of NH topological sensors~\cite{budich_non-hermitian_2020}. Here it is shown that coupling the ends of the anisotropic SSH model, cf. Eq.~\eqref{eq:nh_SSH}, results in an exponential splitting of the eigenvalues, such that perturbing this coupling leads to an exponentially strong response. A crucial ingredient in this context is the presence of a single topological end state, whose right and left eigenvectors localize on opposite ends. This mechanism is exploited in the experimental realization of the NH topological ohmmeter in Ref.~\citenum{konye_non-hermitian_2024}. Other NH sensing protocols exist, such as in Ref.~\citenum{mcdonald_exponentially-enhanced_2020}.

The skin effect has also been related to directional amplification~\cite{wanjura_topological_2020,xue_simple_2021,brunelli_restoration_2023}. Here, a signal entering a NH chain at one end gets exponential amplified towards the other end, while it gets exponential suppressed in the other direction. This principle has been experimentally verified in the bosonic Kitaev chain~\cite{mcdonald_phase-dependent_2018}, a model similar to the Hatano-Nelson model in Eq.~\eqref{eq:HN_Ham}, in a nano-optomechanical network~\cite{slim_optomechanical_2024} and in a quantum simulator in Ref.~\cite{busnaina_quantum_2024}.

\section{Outlook}

Let us end with an outlook. While the NH skin effect is well understood in 1D systems, a complete theoretical framework is needed for higher dimensions, where we discussed the amoeba formalism as one important cornerstone to tackle such problems. Exploring connections to modern mathematics, including K-theory and symmetry classifications, offers promising avenues for uncovering and engineering the NH skin effect in higher-dimensional systems.

At the same time, the interplay between interactions in NH topological phases and skin modes still needs to be fully understood. For example, recent simulations on skin clustering in strongly interacting bosonic systems highlight the fragmented Hilbert space~\cite{zhang_observation_2022} and reveal new connections to phenomena like the eigenstate thermalization hypothesis and quantum scars. Also, going to higher dimensions in correlated fermionic systems opens new avenues to explore many-body skin effects~\cite{yoshida_real-space_2021}.

Beyond what was touched on in the main text, we see great prospects in analyzing the impact of time-periodic driving, i.e., Floquet engineering, in single-particle and interacting many-body systems~\cite{zhou_non-hermitian_2023,faugno_interaction-induced_2022}. Also time dynamics and the steady state nature of NH skin modes in open quantum systems promise new insights~\cite{song_non-hermitian_2019-1}. Finally, we see the introduction of nonlinearities~\cite{yuce_nonlinear_2021} as a great source for novel phenomenology related to the skin effect, leading to modified localization properties, such as trap-skin states~\cite{ezawa_dynamical_2022}, and in general, phenomena beyond established topological theories~\cite{yoshida_exceptional_2025,wang_nonlinear_2024}.

\acknowledgments

The authors would like to thank Anton Montag, Lars Koekenbier, Hermann Schulz-Baldes and Angelo Carollo for insightful discussions, and Federico Roccati and Lukas R{\o}dland for proofreading this text. J.T.G., A.B. and F.K.K. acknowledge funding from the Max Planck Society Lise Meitner Excellence Program~\mbox{2.0}. J.T.G. and F.K.K. also acknowledge support from the European Union’s ERC Starting Grant “NTopQuant” (101116680). The views expressed are those of the authors and do not necessarily reflect those of the European Union or the ERC.

\bibliographystyle{eplbib.bst}
\bibliography{references.bib}
\end{document}